\documentclass[twocolumn,floatfix,showpacs]{revtex4}
\usepackage{graphicx,epsfig,color}

\begin{document}

\title{Electronic and optical properties of Cr and Cr-N doped anatase TiO$_2$ \\ 
from screened Coulomb hybrid calculations}

\author{ Veysel \c{C}elik and  
Ersen Mete\footnote{Corresponding author: \indent e-mail:
emete@balikesir.edu.tr} }

\affiliation{Department of Physics, Bal{\i}kesir University,
Bal{\i}kesir 10145, Turkey}

\date{\today}

\begin{abstract}
We studied the electronic and atomic structures of anatase TiO$_2$ codoped with 
Cr and N using hybrid density functional theory calculations. Nonlocal screened 
Hartree-Fock exchange energy is partially mixed with traditional semilocal 
exchange part. This not only heals the band gap underestimation but also improves 
the description of anion/cation-driven impurity states and magnetization of the 
dopants.  Cr and/or N doping modifies the valence and conduction band edges of 
TiO$_2$ leading to significant band gap reduction. Hence, Cr, N and Cr-N 
doped TiO$_2$ are promising for enhanced photoactivity.
\end{abstract}


\maketitle

\section{Introduction}

For photocatalysis, titanium dioxide (TiO$_2$) offers excellent oxidation and 
charge transport properties together with its wide availability, nontoxicity, and 
chemical stability. For this reason, it finds many useful applications such as 
photogeneration of hydrogen from water, degradation of pollutants under 
visible light irradiation and production of hydrocarbon fuels and dye sensitized 
solar cells (DSSC).\cite{Fujishima,Gratzel,Khan,Varghese} Among the other 
polymorphs the anatase phase exhibits higher catalytic activity.\cite{Xu} 

One important limitation is that anatase TiO$_2$ has a wide band gap of  
$\sim$3.2 eV~\cite{Tang} and can only absorb in the ultraviolet (UV) 
region ($\lambda\!<\,$380 nm). This seriously reduces solar energy 
utilization to $\sim$5\%.  Substitutional cation and/or anion modified 
titania has been proposed as an effective approach to get catalytic activity 
under visible light irradiation.\cite{Wang,HYu,Zhu,Long,Yamamoto,Yin,Celik} 
Recent experiments have shown that Cr-N codoping in TiO$_2$ drastically
enhances absorbance.\cite{Chiodi,McDonnell} Moreover, Cr and Cr-N dopings 
were reported to increase the visible light reactivity with increasing Cr 
incorporation.~\cite{Li,Liu} Ferromagnetism has also been observed in Cr doped 
anatase films.\cite{Osterwalder,Zhang} Theoretical studies predicted magnetization 
per Cr much larger than experimental findings.\cite{Ye,Peng} Pure density functional
theory (DFT) methods unreliably estimate half-metallic character for Cr doped 
TiO$_2$\cite{Peng,DiValentin}. For the optical properties, hybrid theoretical 
approaches has been recently used to get improved quantitative agreement with 
the experimental data.\cite{McDonnell,Kurtoglu} Detailed electronic structure
investigations are still needed to get a proper description of band gap features
for the doped systems.

In this present work, we used screened Coulomb potential hybrid DFT calculations
to investigate the modifications of the band gap properties of TiO$_2$ induced by 
Cr and Cr-N dopants and their effect on the corresponding absorption spectra.
Formation energies have been calculated as a function of oxygen chemical potential 
to compare thermodynamical statbility of the doped structures. We also obtained the 
magnetic moments and the charge states of Cr and N species.

\begin{figure*}[t!]
\epsfig{file=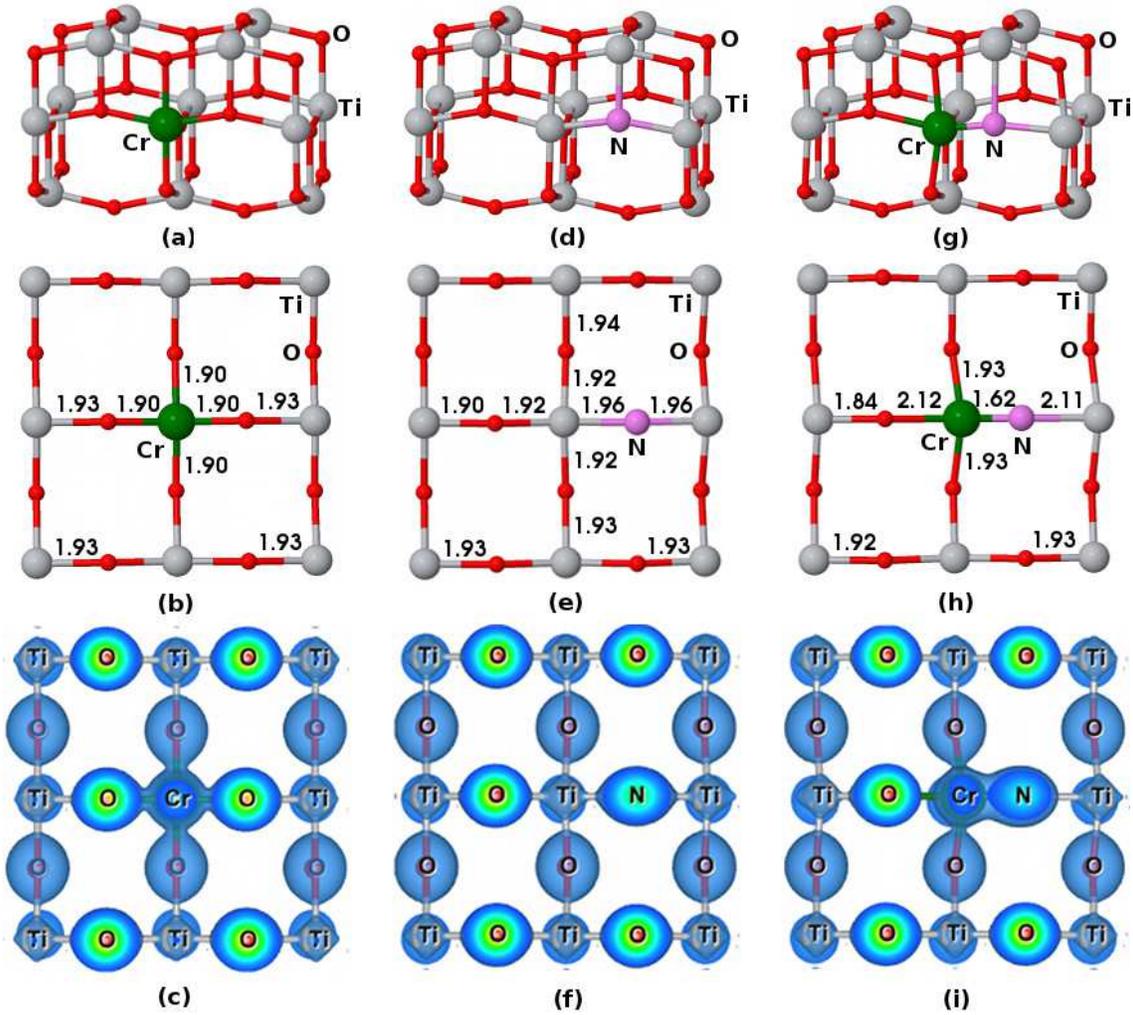,width=15cm}
\caption{Relaxed geometries and total charge densities of 
substitutional N (a-c), Cr  (d-f) and Cr-N (g-i) in anatase TiO$_2$. 
\label{fig1}}
\end{figure*}

\section{Theoretical Method}

For all systems, our calculations were performed based on the spin polarized 
hybrid density functional theory as implemented in the Vienna \textit{ab-initio} 
simulation package (VASP).\cite{vasp} Ionic cores and valence electrons were 
treated by projector-augmented waves (PAW) method.\cite{paw1,paw2} The kinetic 
energy cutoff value was determined to be 400 eV. 

Pure DFT describes impurity and defect associated properties of transition 
metal oxides incorrectly. For example, TiO$_2$ is predicted to be metallic 
in oxygen vacancy cases.\cite{Unal} However, experimental studies show a 
semiconducting nature. This failure of DFT lies in the approximations to 
the exchange-correlation (XC) energy being usually local or semilocal. 
This flaw can be patched by using hybrid methods where a portion of the 
non-local exact exchange is admixed with the traditional semilocal 
exchange. We used the Heyd-Scuseria-Ernzerhof hybrid XC functional 
(HSE06)~\cite{Heyd1,Heyd2,Paier} based on  screened Coulomb potential 
reducing the self-interaction error. Technically, this results in a rapid spatial 
decay of HF exchange which improves the convergence behavior of 
self-consistent iterations. The HSE exchange is derived from the 
PBE0\cite{PBE,Adamo} exchange by range separation and then by cancellation 
of counteracting long range HF and PBE contributions, as, 
\[
E_{x}^{\rm HSE}=aE_{x}^{\rm HF,SR}(\omega)+(1-a)E_{x}^{\rm PBE,SR}(\omega)+E_{x}^{\rm PBE,LR}(\omega),
\]
where $\omega$ is the range separation parameter for the screening and $a$ is 
the mixing coefficient. The parameter of $\omega$=0.2 \AA$^{-1}$ is used 
for the exchange contributions as suggested for the HSE06 functional.\cite{Krukau} 
We determined the mixing ratio of the exact exchange to be 22\% so that calculated 
band gaps and lattice constants for both the anatase and the rutile polymorphs show 
good agreement with the experimental values.\cite{Celik} 

In order to model the doped systems we constructed a 108-atom supercell by 
3$\times$3$\times$1 replication of the conventional anatase unit cell. This
structure is sufficiently large to accommodate spatial separation between the periodic 
images of the defects. We traced all possible Cr and N doping configurations. In 
particular, For Cr/N codoping, the structure shown in Fig.~\ref{fig1} is found to be 
energetically the most favorable. For geometry optimization and density of states 
(DOS) calculations we used 8 special $k$-points to perform the Brillouin zone 
integrations. The calculated properties are converged such that they unnoticeably 
change when a higher $k$-point sampling is used. We required a precision of 
0.015 eV/{\AA} in the residual forces in every spatial component on all the atoms 
without fixing them to their bulk positions. 

The formation energies of the defects were calculated by using the formula,
\[
E_{f}=E_{\rm doped}-E_{\rm pure}-n\mu_{\rm Cr}-m\mu_{\rm N}+n\mu_{\rm Ti}+m\mu_{\rm O},
\]
where E$_{\rm doped}$ (E$_{\rm pure}$) is the total energy of doped 
(pure) supercell. The chemical potentials of Cr, N, Ti and O are
denoted by $\mu_{\rm Cr}$, $\mu_{\rm N}$,  $\mu_{\rm Ti}$ and $\mu_{\rm O}$, 
respectively. Depending on the presence of Cr (N) substitutional dopant, 
the number $n$ ($m$) assumes values 0 or 1. In thermodynamical 
equilibrium, the chemical potentials of Ti and O varies depending on the growth 
environment and must also satisfy the restriction 
$\mu_{\rm TiO_{2}}=\mu_{\rm Ti}+2\mu_{\rm O}$.  
Under O-rich conditions, $\mu_{\rm O}$ is taken from molecular oxygen, and 
then $\mu_{\rm Ti}=\mu_{\rm TiO_{2}}-E_{\rm O_2}$. 
Under Ti-rich conditions, $\mu_{\rm Ti}$ is derived from bulk (hcp) Ti 
($\mu_{\rm Ti}^{\rm bulk}$) and $\mu_{\rm O}$ is calculated from 
$\mu_{\rm O}=\frac{1}{2}(\mu_{\rm TiO_{2}}-\mu_{\rm Ti})$. 
The remaining  chemical potentials of Cr and N are taken from 
their natural phases ($\mu_{\rm Cr}=\frac{1}{2}(E_{\rm Cr_2O_3}-\frac{3}{2}E_{\rm O_2})$ 
 and $\mu_{\rm N}=\frac{1}{2}E_{\rm N_2}$). 

Bader analysis quantitatively describes local charge 
depletion/accumulation. It involves integration of Bader volumes 
around atomic centers. These volumes are partitions of the real space 
delimited by local zero-flux surfaces of charge density gradient vector field.
We calculated charge states of atomic species (in Table~\ref{table1}) 
using a grid based decomposition algorithm.\cite{Henkelman}

\begin{table}[t!]
\caption{Average charge states ($e$) of dopants and their adjecent (nn)
Ti and O atoms from Bader analysis.\label{table1}}
\begin{ruledtabular}
\begin{tabular}{cccccc}
Doping & Cr   & N     & Ti(nn)& O(nn)  \\[1mm]\hline
none   &      &       & +2.84 &$-$1.43 \\
N@O    &      &$-$1.38& +2.31 &        \\
Cr@Ti  & +2.30&       &       &$-$1.33\\
Cr/N   & +2.29&$-$1.13& +2.81 &$-$1.19 \\
\end{tabular}
\end{ruledtabular}
\end{table}

\begin{figure*}[t]
\epsfig{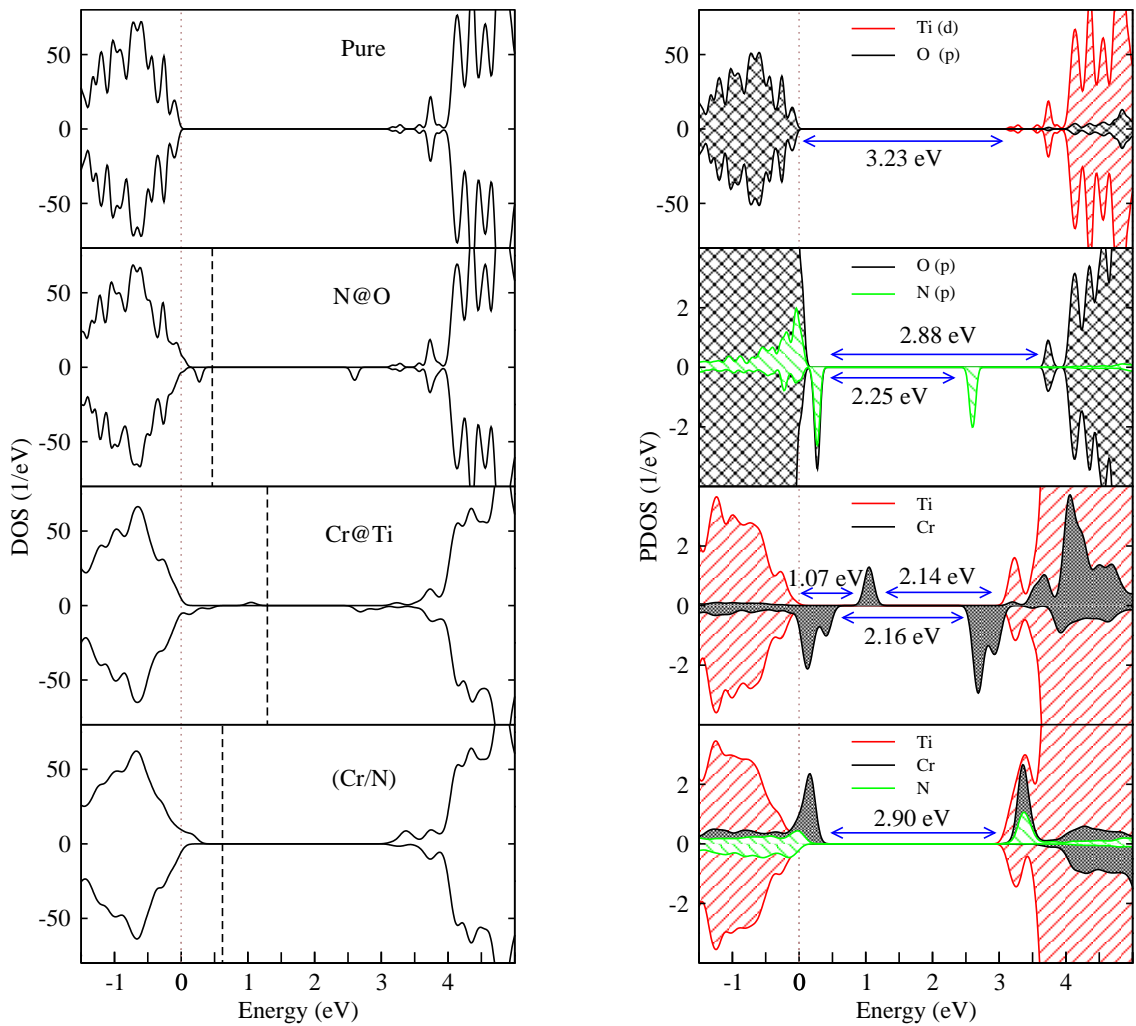}
\caption{Total (left) and projected (right) densities of states (DOS) of
pure and doped (with Cr and/or N) anatase TiO$_2$, calculated with HSE06 
functional. Dashed (dotted) line indicates the Fermi energy (the VBM of pure 
anatase).\label{fig2}}
\end{figure*}

\section{Results \& Discussion}

The bulk properties of pure anatase were calculated using the 108-atom supercell.
Our HSE-predicted lattice parameters ($a\,=\,3.78$ {\AA}, $c\,=\,9.45$ {\AA}) 
and band gap value (3.23 eV in Fig.~\ref{fig2}) show remarkable agreement with 
the experimental data. Since nitrogen doping case has been discussed in detail  
elsewhere,\cite{Celik} we will focus on Cr and Cr-N doped anatase to elucidate 
the role of Cr in photoelectrochemistry of TiO$_2$.

\begin{figure}[t!]
\epsfig{file=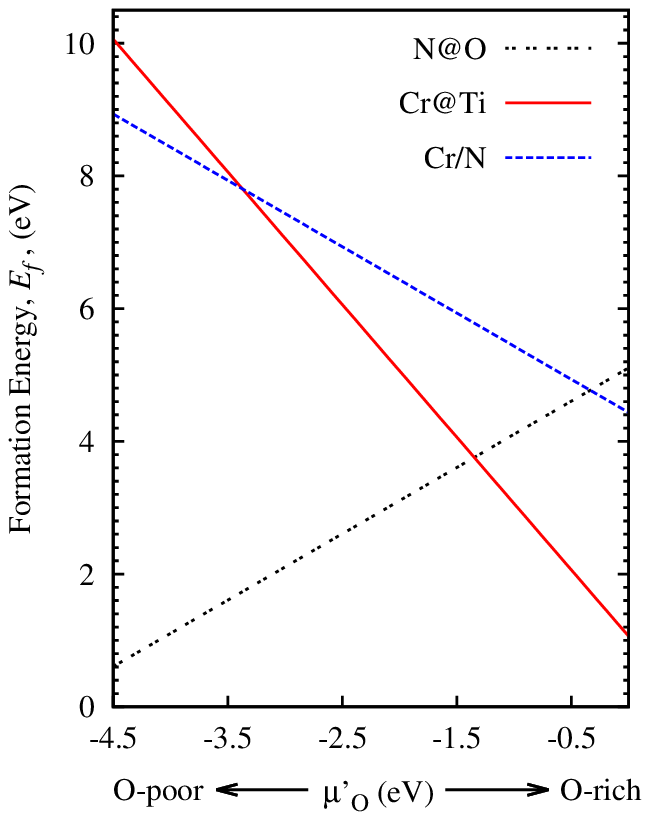,width=7cm}
\caption{Calculated formation energies as a function of the oxygen chemical 
potential (relative to the value at its molecular gas phase)  for Cr, N mono- 
and Cr/N co-doped TiO$_2$ structures..\label{fig3}}
\end{figure}

\vspace{3mm}\noindent\textit{\textbf{Cr-doped TiO$_2$ :}}
Experiments report incorporation of chromium at substitutional sites forming 
single-crystalline Cr-doped TiO$_2$.\cite{Osterwalder,McDonnell,Zhu,Li}  They 
also conclude that if chromium oxide forms it must be highly dispersed and its 
size must be undetectably small.\cite{Chiodi,McDonnell,Li,Liu,Osterwalder,Pan} 
In order to model the structure we substituted one Cr atom at a Ti site in the 
anatase supercell forming sixfold coordination with the nearest neighbor oxygens 
as shown in Fig.~\ref{fig1}. All Cr-O bonds are 1.90 {\AA} and slightly shortened 
relative to those of the undoped TiO$_2$. The disturbance of the substitutional 
Cr on the lattice is very small agreeing with the X-ray diffraction (XRD) 
patterns.\cite{Pan,Liu} Under O-rich conditions, the formation energy of Cr@Ti
doping has been calculated to be 1.06 eV (in Fig.~\ref{fig3}). These energetics 
are sensitive to the choice of the XC functional. For example, McDonnell 
\textit{et al.} found a value of 0.36 eV with HSE06-PBEsol method.\cite{McDonnell}  
Under Ti-rich conditions the formation energy gets as large as 10.21 eV. Therefore, 
synthesis of substitutional Cr under O-rich conditions must be much easier as 
pointed out by previous calculations.\cite{Kurtoglu}

Substitutional Cr introduces $3d$ states which are dominant above the top of 
the valence band (VB) as presented in the corresponding projected DOS (PDOS) of 
Fig.~\ref{fig2}. For the minority spin component, fully occupied Cr $3d$ states 
appear isolated 1.07 eV above the valence band maximum (VBM). Our calculations 
are interestingly in agreement with the X-ray Photoemission Spectra (XPS) of 
Osterwalder \textit{et al.}\cite{Osterwalder} who reported the formation of 
defect states at 1.0 eV above the VBM giving an intensity roughly proportional 
to the Cr concentration. HSE06 functional not only gives the relative positions 
of the defect states in the band gap but also better describes the character of 
these states. For instance, unlike the full-potential linearized augmented plane 
wave (FLAPW) calculations of Ye~\textit{et al.}.\cite{Ye} and LDA study of
Peng~\textit{et al.}, our HSE06 results predict Cr doped anatase to be 
semiconducting, not half-metallic. 

Chromium significantly modifies the conduction band minimum (CBM), too, by 
introducing empty Cr $3d$ states at the bottom of the CB. Widely dispersing
Cr-induced states effectively reduce the calculated band gap from 3.23 eV to 
a value of 2.16(2.14) eV for the spin up(down) component. This band gap 
narrowing agrees well with the decrease of the binding energy of Cr $3d$ from 
3.2 eV to 2.20 eV observed in the XPS spectra.\cite{Osterwalder} Depending 
of the choice of XC flavor, the band gap reduction was calculated by the 
previous theoretical studies to be 1.2 eV,\cite{Ye} and 0.34 eV\cite{McDonnell} 
with FLAPW and HSE06-PBEsol methods, respectively. 

XPS core level and valence band data show that the majority of Cr is present 
in the $3+$ charge state.\cite{Osterwalder,Zhu,Li,McDonnell}  In addition, the 
room temperature magnetization measurements reveal a ferromagnetic state for 
all Cr-doped anatase films, with a saturation magnetic moment of $\sim$0.6 
$\mu_{\rm B}$ per Cr atom.\cite{Osterwalder} Similarly, Zhang~\textit{et al.} 
reported Cr$^{3+}$-associated ferromagnetism which increases up to 
$\sim$0.42 $\mu_{\rm B}$/Cr by lowering of oxygen pressures.\cite{Zhang} 
Theoretical studies, on the other hand, predicted a charge state of $4+$ with 
a magnetization of 2 $\mu_{\rm B}$.\cite{Ye,DiValentin,Kurtoglu} The 
discrepancy has been attributed to the possible presence of oxygen vacancies 
which leave excess charge in the samples reducing the charge state from Cr$^{4+}$ 
to Cr$^{3+}$. However,  without such a compensation, our HSE-calculated 
charge state of  $+2.3$ and total magnetic moment of 1.21 $\mu_{\rm B}$ per Cr 
show a significant improvement over  the existing theoretical estimations.

\begin{figure}[t!]
\epsfig{file=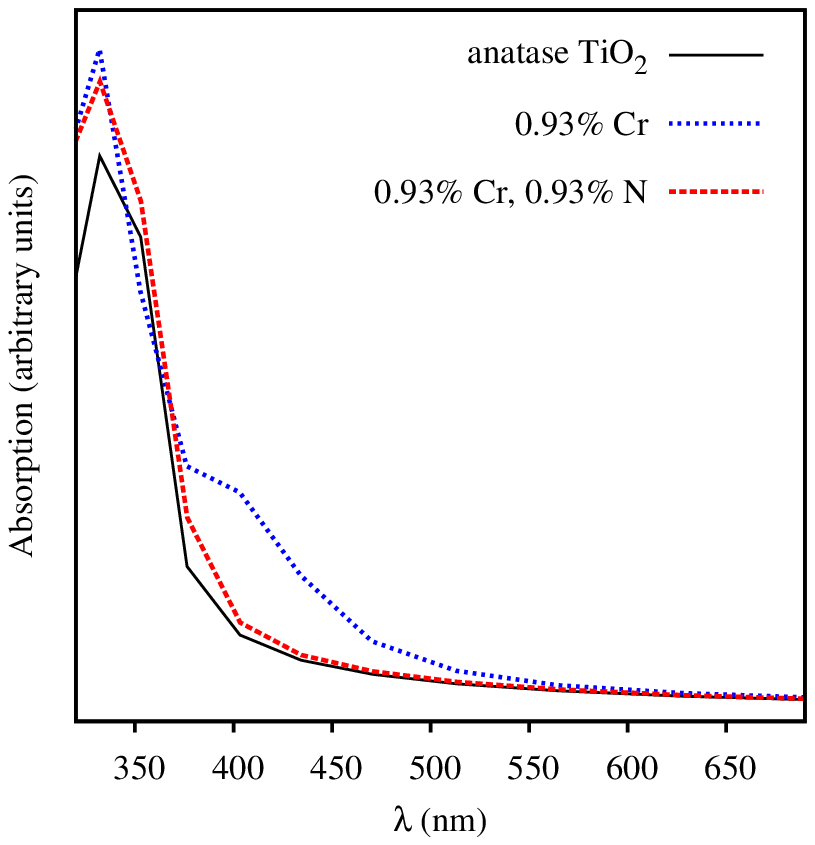,width=8cm}
\caption{Calculated absorption spectra for pure, Cr- and Cr/N-doped anatase TiO$_2$
structures.\label{fig4}}
\end{figure}

\vspace{3mm}\noindent\textbf{\textit{Cr/N-codoped TiO$_2$ :}}
Substitutional codoping of Cr and N in the anatase form has been confirmed by 
XRD measurements for various impurity concentrations.\cite{Chiodi,Li,Liu,McDonnell,Kurtoglu}
Initially, we considered all possible doping configurations. After the relaxation, 
the Cr-N pairing shown in Fig.~\ref{fig1}(g-h) yields the lowest total energy. 
The lattice gets slightly disturbed because of the strong Cr-N interaction leading 
to a bond length of 1.62 {\AA} which is considerably shorter than the Ti-N bonding. 
The total charge density plots in Fig.~\ref{fig1}(g) and Fig.~\ref{fig1}(f) 
clearly shows the covalent character of Cr-N coupling while Ti-N bonding is 
more polarized. 

We calculated the formation energy for Cr/N codoping to be 4.43 eV under O-rich 
conditions. It rises up to 8.93 eV under O-poor growth environment as shown 
in Fig.~\ref{fig3}. The experimental realization of Cr-N codoping is expected 
to be energetically viable under oxidizing atmosphere in agreement with 
previous studies.\cite{McDonnell}

The presence of substitutional N in the codoped anatase phase reduces the 
calculated magnetic moment to 1.00 $\mu_{\rm B}$ relative to the case 
of Cr monodoping. Besides, it insignificantly alters the charge state of Cr 
which is $+2.29$ as presented in Table.~\ref{table1}. This indicates a strong 
delocalization of N $2p$ states over TiO$_2$ bands. The UV-vis reflectance 
spectral measurements of Li \textit{et al.}\cite{Li} and the X-ray absorption 
spectroscopy (XAS) of Chiodi \textit{et al.}\cite{Chiodi} confirm the presence 
of Cr$^{3+}$ species. Our HSE-calculated charge state is reasonably closer 
than the previous DFT+U predicted Mulliken charges on the substitutional 
Cr when N dopant is present.\cite{Kurtoglu}
 
Strong Cr-N pair interaction exhibits interesting features in the electronic
properties in Fig.~\ref{fig2}. First, the isolated impurity states associated with 
Cr $3d$ and N $2p$ in the (Cr@Ti) and (N@O) monodoping cases mutually 
passivate each other giving a clean band gap. This eliminates the possiblility 
of photo-generated charge recombination due to well localized trap states.
Secondly, the Cr and N concentration of $\sim$0.93\% in the computational 
cell causes a band gap reduction of 0.3 eV.  McDonnell \textit{et al.} calculated 
2.66 eV band gap for Cr/N codoping by using HSE06-PBEsol  XC 
functional.\cite{McDonnell} Chiodi \textit{et al.} reported the appearance of
codoping-induced and Cr $3d$ dominant states at the top of the VB which 
translates to a band gap of 2.8 eV.\cite{Chiodi} This observation nicely fits in 
our PDOS structure in Fig.~\ref{fig2} with an estimated band gap of 2.9 eV.
For higher dopant concentrations, experiments report increased gap narrowing
causing significantly red-shifted optical absorption.\cite{Li,McDonnell,Chiodi} 
Thirdly, N $2p$ states which appear in the monodoping case (N@O) in Fig.~\ref{fig2} 
significantly delocalize over the VB and the CB when Cr is present (Cr/N) due 
to strong Cr-N interaction. This PDOS feature shows remarkable agreement with 
the experimental observations.\cite{Chiodi,Li} Therefore, N has a weaker
effect on the optical response than Cr which dominantly modifies the band 
edges of TiO$_2$.

We calculated the absorption spectra of Cr- and Cr/N-doped cases in comparison 
with pure anatase (see Fig.~\ref{fig4}). Cr seems to be absorbing in the visible region 
significantly better than the Cr-N codoping. However, PDOS structure of Cr@Ti 
shows that the overall photocatalytic efficiency might be lower than the Cr/N case 
due to the charge recombination in the presence of localized $3d$ gap states.  
In fact, Cr-N codoped TiO$_2$ exhibits relatively higher catalytic reactivity 
under visible light irradiation.\cite{Li}   

\section{Conclusions}
The electronic, magnetic and optical properties of Cr- and Cr/N-doped anatase 
TiO$_2$ have been investigated by means of screen Coulomb hybrid calculations. 
Hybrid functionals incorporating exact exchange terms not only improve the 
description of defect states but also are promising to predict magnetization 
properties of $d$-band dopants. We have shown the dominant character of  Cr in 
modifying the band edges of anatase. Substitutional Cr with or without N is very 
effective in band gap reduction causing significant red-shift of the absorption 
edge into the visible spectral region. Isolated N $2p$ states get delocalized 
over TiO$_2$ bands due to mutual passivation by Cr $3d$ states giving a clean 
band gap for Cr/N-codoping. Thus, the enhancement of the optical absorption is 
emergent. Moreover, the calculations imply that the substitutional Cr and N pair 
reduces electron-hole recombination rate improving the overall photocatalytic 
performance.

\begin{acknowledgments}
This study was supported in part by T\"{U}B\.{I}TAK, The Scientific and Technological Research 
Council of Turkey (Grant No. 110T394) through the computational resources provided by ULAKB\.{I}M,
Turkish Academic Network \& Information Center.
\end{acknowledgments}

\end{document}